\documentclass{eas}
\usepackage{graphicx}
\begin{document}

\title{Symbiotic stars in the Local Group of Galaxies} 
\author{J. Miko{\l}ajewska}\address{Nicolaus Copernicus Astronomical Center, Warsaw, Poland}
\author{M.M.~Shara}\address{Department of Astrophysics, American Museum of Natural History, New York, USA}
\author{N.~Caldwell}\address{Harvard-Smithsonian Center for Astrophysics, Cambridge, USA}
\author{K.~Drozd\,$^{1}$}
\author{K.~I{\l}kiewicz\,$^{1}$}
\author{D.~Zurek\,$^{2}$}
\begin{abstract}
Preliminary results of the ongoing search for symbiotic binary stars in the Local Group of Galaxies are presented and discussed.
\end{abstract}
\maketitle
\section{Introduction}

Symbiotic stars (SySt) are interacting binaries, in which an evolved giant (either a normal red giant in S-type, or a Mira surrounded by an opaque dust shell in D-type) tranfers mass to hot luminous, and compact (usually a WD) companion.  The interacting stars are embedded in a rich and complex surroundings, including both ionized and neutral regions, accretion/excretion  disks, interacting winds, and jets. Because of their complex structure SySt are important tracers of late evolutionary phases of low- and medium-mass stars, and excellent laboratories to explore interactions and evolution in binary stars.
Finally, the composition of SySt makes some of them a promising "factory" of SN Ia,  independently of the scenario leading to their eruption. 
For the most recent review see e.g. Miko{\l}ajewska (2012).

While about 300 Galactic symbiotics are known (e.g. Belczynski {\em et al.\/} 2000;  Miszalski {\em et al.\/} 2013; Miszalski \& Mikolajewska 2014: Rodriguez-Flores {\em et al.\/} 2014, and references therein), and a few dozen are relatively well studied, their distances (and hence their component luminosities and other distance-related parameters) are poorly determined. This makes comparison with the theoretical models for their interaction and evolution very challenging.

Fortunately, bright SySt have been detected at the Magellanic Clouds (Belczynski {\em et al.\/} 2000; Miszalski {\em et al.\/} 2014, and references therein) and  even in the Local Group of galaxies (Goncalves {\em et al.\/} 2008; Kniazev {\em et al.\/} 2009; Mikolajewska {\em et al.\/} 2014, hereafter MCS14). These discoveries have been, however, in almost all cases serendipitous, and a systematic search for SySt is necessary to provide sample(s) suitable for any statistical analyses.

Here we present some preliminary results of our ongoing systematic search for SySt in M33 and M31. 

\section{Selection method}

The motivation and the basic concepts for this project have been presented by MCS14. In particular, we expect to obtain large and complete luminosity-limited samples of extragalactic SySt to estimate their total numbers, their spatial distribution, and their basic physical parameters. 

The candidates have been selected based on photometric measurements on the publicly available survey LGGS images (Massey {\em et al.\/} 2007). First, we have created catalogs of targets with the following criteria:
(i) visible at least in $VRI$  and H$\alpha$, (ii) red, i.e. $V-I \geq 1$, and (iii)  H$\alpha$-$R \leq 0$. In addition, we rejected objects with $I > 21$ (which corresponds to the faintest SySt detected in M31 by MCS14). The resulting catalogs contained over 500,000 objects in M31 and M33.
SySt should be simultaneously red (due to the presence of a cool giant) and show strong emission lines (from the ionized nebula). 
So, as possible SySt candidates we have selected those with $V-I \geq 1.5$, and 
$-1 \geq {\rm H}\alpha-R \geq -4$, based on $VI$H$\alpha$ magnitudes of SySt from MCS14. These have given us $\sim 2000$ candidates in M31, and another $\sim 800$ in M33.

\section{Preliminary results from follow-up spectroscopy}

To confirm which of these objects are SySt, we obtained spectra of $\sim 1300$ of them using the Hectospec multi-fiber positioner and spectrograph on the 6.5m MMT telescope (details in MCS14, and references therein).

\begin{figure}
\includegraphics[width=0.46\textwidth]{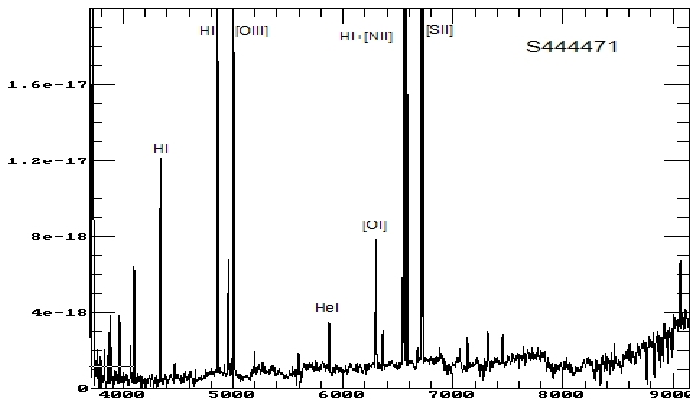}
\qquad
\includegraphics[width=0.46\textwidth]{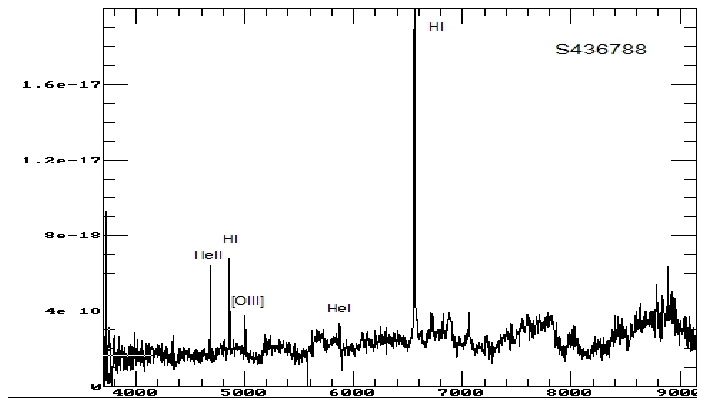}
\caption{Spectra of two SySt candidates in M33, both showing strong absorption features of a carbon giant and strong emission lines. However, only the spectrum on the right can be attributed to SySt according to our criteria. The emission lines in the left spectrum are from DIG (see text).} 
\label{F1}
\end{figure}

To classify an object as SySt, most authors adopt Kenyon's (1986)  definition, which, in addition to the presence of absorption features of a late-type giant, requires strong H\,I and He\,I emission lines, and additional lines with an ionizational potential of at least 30 eV (e.g. [O\,III]) and an equivalent width $ > 1\,\rm \AA$.  
Most of the spectra revealed the presence of a red giant and strong emission lines satisfying this definition (Fig.~\ref{F1}). However, the forbidden [O\,II], [O\,III], [N\,II] and [S\,II] line ratios are consistent with low-density ($n_{\rm e} \sim 100\, \rm cm^{-3}$ or less) formation regions, and they do originate in diffuse ionized gas (DIG) in M33 and M31 rather than in the much denser symbiotic nebula. Such DIG is present in all LGG disks, and is particularly abundant in star forming regions (e.g. Hoopes \& Walterbos 2003), and it may significantly pollute the SySt candidate spectra. 

Therefore, we have decided to use stronger criteria to accept a star as SySt, in particular, the presence of He\,II and higher ionization emission lines (Fig.~\ref{F1}).
Unfortunately, these have given us only a dozen SySt (out of almost 500 candidates observed) in M33, and about 50 new SySt (out of about 750 observed) in M31 (Miko{\l}ajewska {\em et al.\/} 2015, in preparation).
Additional criteria have been applied to identify possible SySt. These involve the [O\,III] and He\,I diagnostic diagrams, which allow to distinguish between dense SySt nebulae and lower density planetary nebulae and H\,II regions (for details see MCS14, and references therein).
Another category of possible low-ionization SySt are objects with cool giant absorption spectra and strong H\,I emission lines, and no signature of the low-density DIG lines. As a result we found a few possible SySt in M33, and about 20 in M31, however, the analysis of the spectra has not yet been completed, and more may turn up.
Even now, our SySt in M33 and M31 are the largest extragalactic symbiotic samples; there are only 16 SySt detected so far in both Magellanic Clouds, and only a few in other LGG. Moreover, distances to the Galactic SySt are poorly determined, and in most cases their accuracy is only 50\%, or less. 
While our sample sizes are modest, the uniquely well-determined distances and luminosities allow us to present some preliminary but important conclusions.

First of all, there is very striking difference between the M33 and M31 SySt. 
In M33, eight SySt have an M-type giant, but in four of them a carbon giant is present. In M31, the situation is radically different, because only one strong SySt and two possible SySt among almost 100 (including SySt from MCS14) contain C-rich giant. A similarly high ratio of C/M giants is observed in the MC SySt (C/M$\sim$2 and 0.4 in LMC and SMC, respectively), whereas a low C/M$\sim$0.05 characterizes those in the Milky Way (MW). Since it is generally believed that the C/M giant ratio is related to metallicity of the parent population, the high fraction of C-giants in M33 and MC SySt is likely reflecting their low metallicity. 
Similar C/M ratios result from the not-symbiotic giants in our sample: C/M$\sim$0.3 and $\sim$0.02 in M33 and M31, respectively.      

\begin{figure}
\centerline{\includegraphics[width=0.8\textwidth]{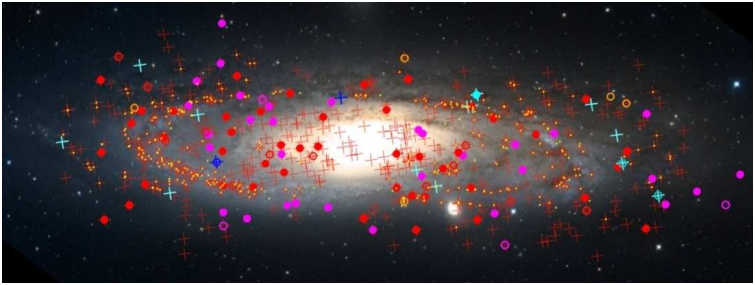}}
\caption{Distribution of the observed objects overlaid on the DSS optical image of M31: crosses -- objects with red giants (M-type -- red, C-type -- cyan, and blue -- S star); circles -- new SySt (colors reflect the giant type) and SySt from MCS14 (fuchsia), where full and open symbols are SySt - with He\,II emission, and possible (based on H$\alpha$ -- red and [O\,III] diagram -- orange) SySt, respectively; yellow dots --  objects with DIG emission.}
\label{F2}
\end{figure}

Fig.~\ref{F2} shows the distribution of so far observed objects together with SySt in M31.
SySt are distributed over the whole disk up to $\sim 25$ kpc from the galactic center, and $\sim 40\,\% $
of them are in the outer disk, beyond 15 kpc from the center. This is  a remarkable result because of the scarcity of SySt currently known in MW beyond the Sun's orbit. Fig.~\ref{F2} indicates that significant numbers of anticenter Galactic SySt remain to be found.

23 of our SySt coincide with red variables (not Miras) found in the  POINT-AGAPE Survey of M31 (An {\em et al.\/} 2004). Their periods fall in the range covered by the orbital periods of known SySt, and the period distribution peaks between 600 and 1000 days like that of MW SySt (Mikolajewska 2012).

Some serendipitous discoveries have been also made. Among the most interesting is the finding that the most luminous infrared star in M33 is almost certainly a binary composed of a massive O star and a dust-embedded red hypergiant (Miko{\l}ajewska {\em et al.\/} 2015). We have also found the first (and only one) WCE/WNE transition star in M31, 
displaying both hot WC and WN Wolf-Rayet emission lines (Shara {\em et al.\/} 2015, in preparation).

\noindent {\bf Acknowledgement.} This study has been supported in part by the Polish NCN grant DEC-2013/10/M/ST9/00086.


\end{document}